\documentclass[usenatbib]{mn2e}
\usepackage{graphicx,ifthen,url,float}
\newcommand {\um}{$\mu$m}

\def\um     {$\mu$m}
\def\ts     {\thinspace}
\def\kms    {\ifmmode{{\rm \ts km\ts s}^{-1}}\else{\ts km\ts s$^{-1}$}\fi}
\def\msol   {\ifmmode{{\rm M}_{\odot}}\else{M$_{\odot}$}\fi}
\def\lsol   {\ifmmode{{\rm L}_{\odot}}\else{L$_{\odot}$}\fi}
\def\zsol   {\ifmmode{{\rm Z}_{\odot}}\else{Z$_{\odot}$}\fi}
\def\etal   {{\rm et\ts al.}}
\def\aco    {\ifmmode{^{12}{\rm CO}(J\!=\!1\! \to \!0)}\else{$^{12}${\rm CO}($J$=1$\to$0)}\fi}
\def\bco    {\ifmmode{^{12}{\rm CO}(J\!=\!2\! \to \!1)}\else{$^{12}${\rm CO}($J$=2$\to$1)}\fi}
\def\cco    {\ifmmode{^{12}{\rm CO}(J\!=\!3\! \to \!2)}\else{$^{12}${\rm CO}($J$=3$\to$2)}\fi}
\def\dco    {\ifmmode{^{12}{\rm CO}(J\!=\!4\! \to \!3)}\else{$^{12}${\rm CO}($J$=4$\to$3)}\fi}
\def\gco    {\ifmmode{^{12}{\rm CO}(J\!=\!7\! \to \!6)}\else{$^{12}${\rm CO}($J$=7$\to$6)}\fi}

\def\ci     {\ifmmode{{\rm C}{\rm \small I}}\else{C\ts {\scriptsize I}}\fi}
\def\hi     {\ifmmode{{\rm H}{\rm \small I}}\else{H\ts {\scriptsize I}}\fi}
\def\hh     {\ifmmode{{\rm H}_2}\else{H$_2$}\fi}
\def\cone {\ifmmode{{\rm C}{\rm \small I}(^3\!P_1\!\to^3\!P_0)}
     \else{C\ts {\scriptsize I}{\small$(^3\!P_1\!\to\,^3\!P_0)$}}\fi}
\def\ctwo {\ifmmode{{\rm C}{\rm \small I}(^3\!P_2\!\to\,^3\!P_1)}
     \else{C\ts {\scriptsize I}{\small$(^3\!P_2\!\to\,^3\!P_1)$}}\fi}
\def\cij {\ifmmode{{\rm C}{\rm \small I}\,(^3P_i\to^3P_j)}\else{C\ts {\scriptsize I}\,{\small$(^3P_i\to^3P_j)$}}\fi}
\def\cii    {\ifmmode{{\rm C}{\rm \small II}}\else{C\ts {\scriptsize II}}\fi}
\def\tex {\ifmmode{{T}_{\rm ex}}\else{$T_{\rm ex}$}\fi}
\def\tmb {\ifmmode{{T}_{\rm mb}}\else{$T_{\rm mb}$}\fi}
\def\tkin {\ifmmode{{T}_{\rm kin}}\else{$T_{\rm kin}$}\fi}
\def\microns {\ifmmode{\mu{\rm m}}\else{$\mu$m}\fi}
\def\nhh   {\ifmmode{n({\rm H}_2)}\else{$n$(H$_2$)}\fi}

\newcommand{\ltaraw}{$\; \buildrel < \over \sim \;$}
\newcommand{\lta}{\lower.5ex\hbox{\ltaraw}}
\newcommand{\gtaraw}{$\; \buildrel > \over \sim \;$}
\newcommand{\gta}{\lower.5ex\hbox{\gtaraw}}

\newcommand{\lya}{{\rm\,Ly-$\alpha$}}

\loadboldmathitalic 
\title [Far-Infrared SED fitting]
{Far-infrared Spectral Energy Distribution Fitting for Galaxies Near and Far}
\author[C.~M. Casey]
{
Caitlin~M. Casey$^1$\thanks{Hubble Fellow; cmcasey@ifa.hawaii.edu}\\
$^1$ Institute for Astronomy, University of Hawai'i, 2680 Woodlawn Dr, Honolulu, HI 96822, USA \\
}
\date{Accepted 2012 June 7.  Received 2012 May 22; in original form 2012 April 9.}
\pagerange{\pageref{firstpage}--\pageref{lastpage}} \pubyear{2012}
\begin{document} 
\maketitle 

\label{firstpage}
\begin{abstract}
Spectral Energy Distribution (SED) fitting in the far-infrared (FIR)
is greatly limited by a dearth of data and an excess of free
parameters$-$from galaxies' dust composition, temperature, mass,
orientation, opacity, to heating from Active Galactic Nuclei (AGN).
This paper presents a simple FIR SED fitting technique joining a
modified, single dust temperature greybody, representing the
reprocessed starburst emission in the whole galaxy,
 to a mid-infrared powerlaw, which approximates hot-dust emission from
 AGN heating or clumpy, hot starbursting regions.  This FIR SED can be
 used to measure infrared luminosities, dust temperatures and dust
 masses for both local and high-$z$ galaxies with 3 to 10+ FIR
 photometric measurements.  While the fitting technique does not model
 emission from polycyclic aromatic hydrocarbons (PAHs) in the
 mid-infrared, the impact of PAH features on integrated FIR properties
 is negligible when compared to the bulk emission at longer
 wavelengths.

This fitting method is compared to infrared template SEDs in the
literature using photometric data on 65 local luminous and
ultraluminous infrared galaxies, (U)LIRGs.  Despite relying only on
2--4 free parameters, the coupled greybody/powerlaw SED fitting
described here produces better fits to photometric measurements than
best-fit literature template SEDs (with residuals a factor of $\sim$2
lower).  A mean emissivity index of $\beta$=1.60$\pm$0.38 and
mid-infrared powerlaw slope of $\alpha$=2.0$\pm$0.5 is measured; the
former agrees with the widely presumed emissivity index of $\beta$=1.5
and the latter is indicative of an optically-thin dust medium with a
shallow radial density profile, $\approx$r$^{-1/2}$.
Adopting characteristic dust temperature as the inverse wavelength
where the SED peaks, dust temperatures $\sim$25--45\,K are measured
for local (U)LIRGs, $\sim$5--15\,K colder than previous estimates
using only simple greybodies.  This comparative study highlights the
impact of SED fitting assumptions on the measurement of physical
properties such as infrared luminosity (and thereby infrared-based
star formation rate), dust temperature and dust mass, for both local
and high-redshift galaxies.
\end{abstract}
\begin{keywords} 
galaxies: evolution $-$ galaxies: high-redshift $-$ galaxies: infrared $-$ 
galaxies: starbursts 
\end{keywords} 

\section{Introduction}\label{introduction}

Modeling galaxies' multiwavelength emission
has become a sophisticated effort of extragalactic astronomy.
Spectral energy distribution (SED) templates, generated by modeling
galaxies' stellar populations and radiation, are used prolifically
to derive stellar masses, extinction corrections, and stellar ages
using broadband photometry from the rest-frame ultraviolet to infrared
wavelengths.  They are also commonly used to constrain redshifts
photometrically \citep[e.g.][]{bolzonella00a}.  The population
synthesis-generated SEDs used to fit short wavelength data
($\lambda\,\le$\,8\um) are complex \citep{bruzual03a,maraston05a}.  They
depend on the initial mass function
\citep[IMF;][]{salpeter55a,kroupa01a,chabrier03a}, metallicity,
stellar age, and starbursting timescale $-$ duration, frequency, and
strength.  Although this gives rise to many free parameters in the
models, the slew of broadband filters in the optical and near-infrared
make this detailed SED fitting possible, even with the effects of dust
obscuration/attenuation taken into account
\citep[e.g.][]{calzetti94a,calzetti01a}.  Follow-up spectral
observations in the optical and near-IR often confirm good fits to
broadband photometry and accurate stellar population modeling.

The dawn of new infrared observing facilities$-$from the {\it Herschel
  Space Observatory}, the Atacama Large Millimeter Array (ALMA), to
the {\sc Scuba2} instrument on the James Clerk Maxwell Telescope
(JCMT)$-$has triggered a wave of interest in extending the use of
these template SED libraries to the far-infrared (FIR;
$\sim$8-1000\um\ rest-frame), for example, in generating far-infrared
photometric redshift estimates \citep{roseboom12a}.  Unfortunately,
modeling the dust emission from galaxies is just as complex in the
far-infrared as it is in the optical since free parameters include
dust composition, dust grain type, galaxy structure, orientation,
active galactic nuclei (AGN) heating, emissivity, and optical depth.
However, in contrast to the optical and near-infrared, far-infrared
observations are plagued with a dearth of data.  Where there might be
10-20 broadbands in the optical/near-IR, there are at most $\sim$8
bands in the FIR (most galaxies having data in 3 bands or less), all
of which suffer from the increased beamsize of single-dish FIR
observations, thus increasing uncertainty on measured flux.

Nevertheless, detailed radiative transfer models and empirical
template libraries have been devised in recent years to model the dust
infrared emission of stars, molecular clouds and starburst galaxies
over a wide range of bolometric luminosities
\citep{silva98a,chary01a,dale01a,dale02a,abel02a,siebenmorgen07a,draine07a}.
These models are then often used as a basis on which to measure the
fundamental parameters of observed starburst galaxies with constrained
photometric measurements, both at low redshift
\citep[e.g.][]{rieke09a,armus09a,chapin09a,u12a} and at high redshift
\citep[e.g.][]{blain02a,chapman04a,pope08a,swinbank10a}. Particularly
for galaxies with high star formation rates, the `UV chimney' argument
\citep{neufeld91a} can be used to relate the modeled dust and
molecular cloud structure back to the \lya\ escape fraction;
\lya\ suffers from less attenuation than UV continuum photons despite
very large reservoirs of dust and gas.

This paper investigates the use of these far-infrared template SEDs to
determine fundamental galaxy properties such as infrared luminosity,
$L_{\rm IR}$, characteristic dust temperature, $T_{\rm dust}$, and
dust mass, $M_{\rm dust}$.  Section \ref{s:seds} presents a simple
method for representing galaxies' FIR emission as a coupled modified
greybody plus a mid-infrared (MIR) powerlaw, which can constrain these
fundamental IR-derived properties quite well for a wide range of
galaxies.  Section \ref{s:compare} compares these fits to the SED
template fits of \citet{chary01a}, \citet{dale02a} and
\citet{siebenmorgen07a}, using the photometry of local luminous
infrared galaxies and ultraluminous infrared galaxies (LIRGs/ULIRGs)
from \citet{u12a}.  Section \ref{s:derived} discusses derived
quantities of FIR SED fits, and section \ref{s:conclude}
concludes. Throughout, a $\Lambda$ {\sc CDM} cosmology with $H_{\rm
  0}$=71\,km\,s$^{-1}$\,Mpc$^{-1}$ and $\Omega_{\rm m}$=0.27 is
assumed \citep{hinshaw09a}.
\begin{figure*}
\centering
\includegraphics[width=1.8\columnwidth]{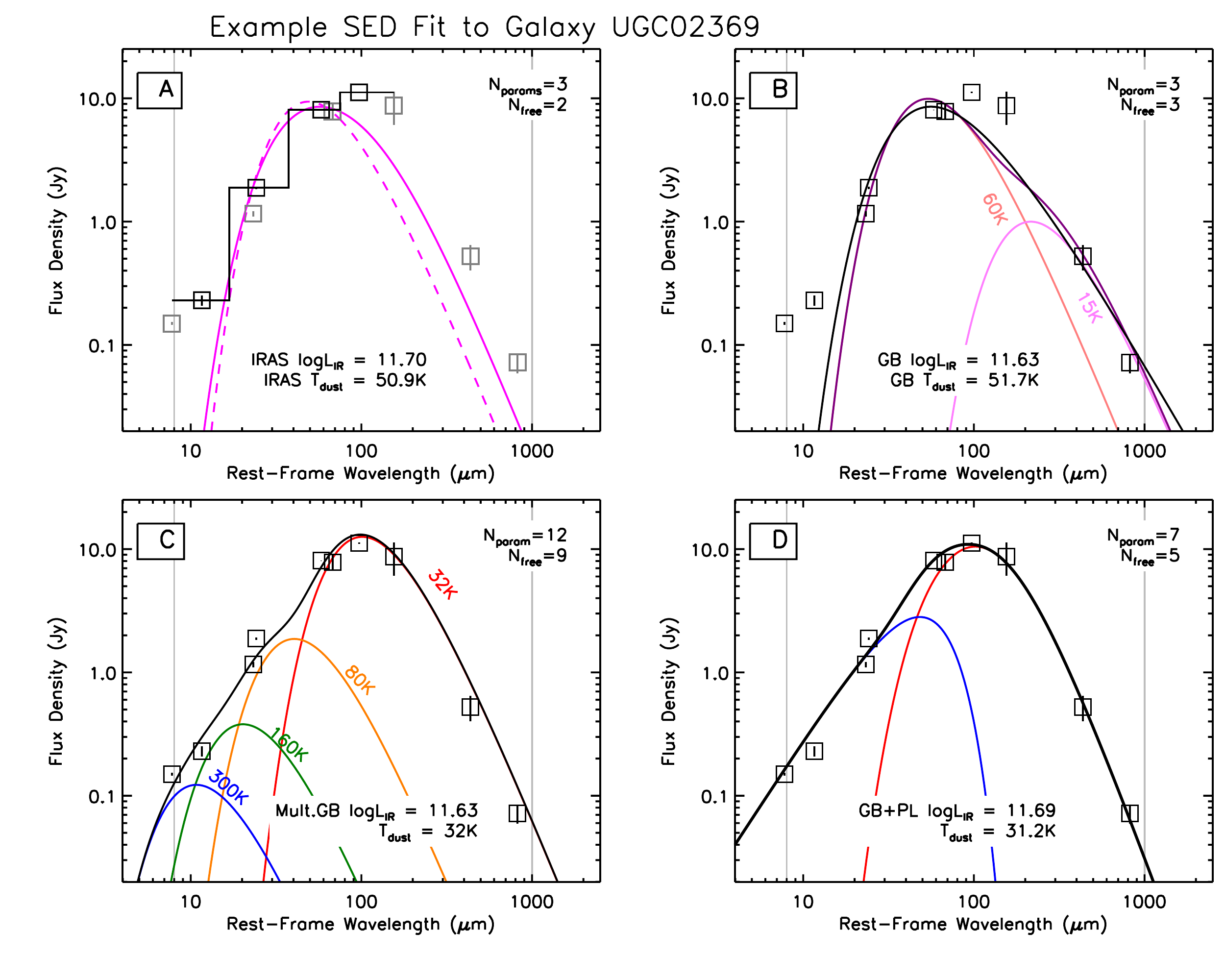}
\caption{Illustration of FIR SED fitting techniques using the example
  local galaxy UGC02369.  Panel (A) highlights the four {\it
    IRAS}-band photometric points in black (the more recent FIR
  photometric measurements in gray).  The original computed FIR
  luminosity for this source was computed via Eq~\ref{eq6}.  The
  original dust temperature for the source (Perault\,1987) was found
  by fitting a single temperature greybody with fixed $\beta$=1.5 to
  the {\it IRAS} bands (magenta; the same best-fit SED assuming
  optically thin conditions is shown in dashed magenta).  Panel (B)
  shows a slightly improved SED fit, with a single temperature
  greybody fit to all of the FIR photometric bands.  Emissivity,
  $\beta$, is added as a free parameter (black) and measured as
  $\beta=$0.5, which is similar to adding a very cold greybody
  component to the warm component dominating the emission, both with
  fixed $\beta\equiv$1.5 (e.g. 60\,K and 15\,K SED shown).  Both the
  fits in panels (A) and (B) have mid-infrared excesses and are poor
  fits to a single temperature greybody.  Panel (C) shows an improved
  fit, using four temperature greybodies with fixed $\beta$=1.5.  The
  fit to the data is much better than (A) or (B), although the number
  of free parameters shoots up to 9.  Panel (D) illustrates the SED
  fitting technique described in \S~\ref{s:sedfitting}, a composite
  mid-infrared powerlaw plus cold-dust single temperature greybody.
  The mid-infrared powerlaw is a good approximation for the composite
  of warm dust giving rise to the mid-IR excess. All fits here assume
  general opacity conditions; in the optically thin case, the fits
  alter slightly, but produce the same $L_{\rm IR}$ and $T_{\rm dust}$
  to within $\sim$1\%. }
\label{fig:schematic}
\end{figure*}

\begin{figure*}
\centering
\includegraphics[width=1.8\columnwidth]{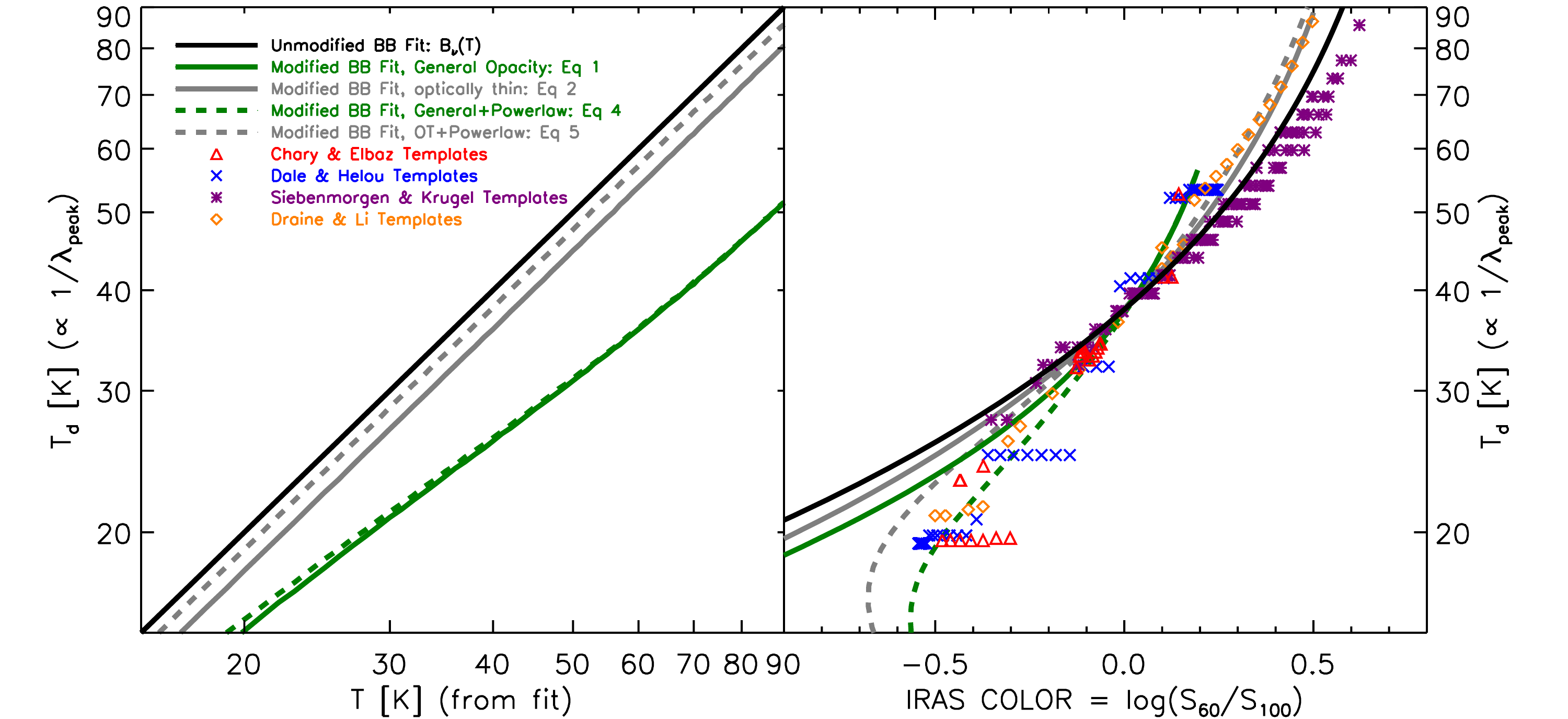}
\caption{ At left, a comparison of the $T$ temperature parameter of
  the fitted greybody against the ``peak wavelength''
  T$_{\rm d}$, which is proportional to the inverse peak in S$_{\nu}$.
  This varies by fitting method mostly due to slight variations in
  adopted opacity, which shifts the SED peak slightly while
  keeping $L_{\rm IR}$ roughly constant.  At right, a comparison
  $T_{\rm d}$, the ``peak wavelength'' temperature, to {\it IRAS}
  [60]-[100] color, which has often been used as a proxy for dust
  temperature.  Over-plotted are various model template values, from
  \citet*{chary01a}, \citet*{dale02a}, \citet*{siebenmorgen07a} and
  \citet*{draine07a}.}
\label{fig:tdlpeak}
\end{figure*}

\section{SED Fitting Techniques}\label{s:seds}

\subsection{Coupled Greybody/Powerlaw Fitting}\label{s:sedfitting}

\subsubsection{Method}

Deriving the fundamental physical properties of IR-luminous galaxies
can be as simple as assuming an isotropically emitting blackbody.
This is represented as the Planck function, $B_{\rm \nu}$($T$)
(e.g. in units of erg\,s$^{-1}$\,cm$^{-2}$\,\AA$^{-1}$), and is only
dependent on dust temperature $T$.  However, if the variation in
opacity (e.g. assuming a screen of dust without scattering) and source
emissivity is accounted for (the fact that very few sources are
perfectly non-reflective), the flux density at rest-frame frequency
$\nu$ is then represented by a modified black body (i.e. ``greybody'')
of the form
\begin{equation}
S({\nu}) \propto (1 - e^{-\tau(\nu)})B_{\nu}(T) = \frac{(1-e^{-\tau(\nu)})\nu^{3}}{e^{h\nu/kT}-1}
\label{eq1}
\end{equation}
where $S(\nu)$ is in units of erg\,s$^{-1}$\,cm$^{-2}$\,Hz$^{-1}$ or
Jy.  Optical depth is $\tau$($\nu$) and fitted as $\tau(\nu) =
(\nu/\nu_{0})^\beta$, where $\nu_{0}$ is the frequency where optical
depth equals unity \citep{draine06a} and $\beta$ represents
emissivity, or the spectral emissivity index.  See \citet{kovacs10a}
for a thorough discussion of the impact on $\beta$.  The value of
$\beta$ is largely assumed to be 1.5 \citep[and usually ranges
  1--2,][]{hildebrand83a}, although this could be a result of the
original wavelengths for which data were gathered on local
starbursting samples \citep{dunne01a}.  Some recent work points to a
wider range of $\beta$ values between 1-2.5
\citep[e.g.][]{chapin11a,casey11a}.  The theoretically expected value
of $\nu_{0}$ is 3\,THz (i.e. $\lambda_{0}$=100\,\um), although this
value is unconstrained by data \citep[see discussion in][]{conley11a}.
In the optically thin case, the term $(1 - e^{-\tau(\nu)})$ reduces to
$\nu^\beta$, and the flux density simplifies to
\begin{equation}
S_{\rm ot}({\nu}) \propto \nu^\beta B_{\nu}(T) = \frac{\nu^{\beta+3}}{e^{h\nu/kT}-1}\,.
\label{eq3}
\end{equation}

The normal range of dust temperatures expected for a galaxy's
interstellar medium (ISM) heated only by star formation ranges
$\sim$20-60\,K.  When fitted to a greybody (as in Equations~\ref{eq1}
or \ref{eq3}), most galaxies have a notable flux density excess at
wavelengths shortward of $\approx$50\um\ (see
Figure~\ref{fig:schematic}, panels A and B).  This mid-infrared excess
is due to a combination of hotter-dust subcomponents (where dust is
more compact) or dust heated by an AGN, and an optically thin medium
by which the higher frequency radiation can escape.  The disconnect
between observed mid-infrared luminosities and predicted Wein-tail
luminosities has been studied for quite some time; a quite thorough
discussion of dust clouds' opacity, radial density distributions, and
dust mass coefficients ($\kappa_\nu$), impact on observed SED is given
in \citet*{scoville76a}; note in particular the difference between
optically thin and optically thick models in the near- and
mid-infrared regime.

Non-thermal emission from polycyclic aromatic hydrocarbons (PAH) in
starburst galaxies or silicate absorption at 9.7\um\ can contribute
to the SED shape in this regime as well; however, their net effect on
integrated IR luminosity, differing from a simple mid-infrared
powerlaw, is $<$10\%\ in most cases (See \S~\ref{s:templates}).

While the hot-dust component might be made up of several subcomponents
of different warm temperatures, the cold dust modified greybody still
dominates the bulk of the total infrared emission when integrated.
The net sum is an SED which can be approximated as a powerlaw in the
mid-infrared, with intensity dropping with decreasing wavelength, and
a single temperature greybody fit in the far-infrared.  The greybody
dominates at wavelengths $>$50\um\ whereas the mid-infrared powerlaw
dominates at wavelengths $<$50\um.  This can be analytically
approximated as:
\begin{equation}
S(\lambda) =
N_{\rm bb}\,\frac{(1-e^{-(\frac{\lambda_{0}}{\lambda})^{\beta}})(\frac{c}{\lambda})^{3}}{e^{hc/\lambda\!kT}-1}
+
N_{\rm pl}\,\lambda^{\alpha}\,e^{-(\frac{\lambda}{\lambda_{c}})^2}
\label{eq4}
\end{equation}
where S$(\lambda)$ is in units of Jy, $T$ is the galaxy's
characteristic ``cold'' dust temperature (in other words, the dust
temperature dominating most of the infrared luminosity and dust mass),
$\lambda_{0}$ is the wavelength at which optical depth is unity
\citep[taken here to be fixed at $\lambda_{0}$=200\um\ as
  in][]{conley11a}, $\beta$ represents the emissivity, $\alpha$
represents the slope of the mid-infrared powerlaw component, and
$\lambda_{c}$ is the wavelength where the mid-infrared 
powerlaw turns over and no longer dominates the emission.  This simplifies to
\begin{equation}
S_{\rm ot}(\lambda) =
N_{\rm bb,ot}\frac{(\frac{c}{\lambda})^{\beta + 3}}{e^{hc/\lambda\!kT}-1}
+
N_{\rm pl}\,\lambda^{\alpha}\,e^{-(\frac{\lambda}{\lambda_{c}})^2}
\label{eq5}
\end{equation}
in the optically thin case.  Note that several works have recognized
the utility of adding a mid-infrared powerlaw components to simple
greybody fits \citep[e.g.][]{younger09a} but tend to fit the two
components separately: first the greybody followed by the mid-infrared
powerlaw.  Coupling the two together and fitting simultaneously makes
it possible for a more accurate fit to be made to systems with fewer
FIR photometric datapoints.

Figure~\ref{fig:schematic} illustrates this SED fitting technique
relative to single temperature greybody fits (as in Eq~\ref{eq1}) for
an example local LIRG, UGC\,02369.  This highlights the inconsistency
of single temperature greybodies with FIR photometry of real galaxies
and the usefulness of an SED fitting technique which simultaneously
fits the cold-dust, long-wavelength greybody and mid-infrared excess
powerlaw component.

Another common way to consider the 'warm dust' components is as a
continuous powerlaw distribution of temperature components,
e.g. $dM_{\rm d}/dT\propto\,T^{-\gamma}$, as described in
\citet{kovacs10a}.  The index $\gamma$ effectively represents
$\alpha$; the values $\gamma$=7.2 and $\gamma$=6.7 \citep[both assumed
  in the ][analysis for different samples]{kovacs10a} correspond to
$\alpha$ values of $\approx$2.9 and 2.6 respectively.
\citet{kovacs10a} points out that $\gamma$$\approx$6.5-7.5 is expected
for sources in a diffuse, star forming medium, and
$\gamma$$\approx$4-5 for a dense medium from the heating source
(e.g. a dusty torus around an AGN).  The flux density function is then
given as
$S_{\nu(T_{c}\rightarrow\infty)}(T)=(\gamma-1)T^{\gamma-1}\int_{T}^{\infty}S_{\nu,T}(T)T^{-\gamma}dT$.
This integral must be solved numerically, but leads to an SED of shape
similar to those produced by Equations~\ref{eq4} and~\ref{eq5},
particularly Equation~\ref{eq4}.  The differences between the
temperature powerlaw integral fit and the analytic approximation in
Eq~\ref{eq4} is that the latter provides a bit more flexibility in
that the relationship between normalization factors can be adjusted.
The former has a clear, robust motivation, but is also a
computationally expensive algorithm.  As photometry for infrared
sources improve and spectroscopy becomes more widely available, a
minority of galaxies will likely have unusual SED shapes not well
described by $dM_{\rm d}/dT\propto\,T^{-\gamma}$, in which case a more
general form of Eq~\ref{eq4} might be appropriate.

Note that another popular infrared SED fitting method joins a greybody
to a mid-infrared powerlaw using a piece-wise technique
\citep{blain02a,blain03a}, where the transition point between greybody
and powerlaw is defined by $d/d\lambda\,(S)$=$\alpha$, i.e. the
gradients are equal.  This is the most straightforward way of
generating SEDs of the desired shape (panel D of
Figure~\ref{fig:schematic}).  In many instances, this method might be
preferred over this paper's fitting techniques (e.g. when generating
hypothetical SEDs for testing selection, or generating fits for
galaxies with only one FIR photometric point).  However, one
disadvantage of this method is its piece-wise nature, making it
difficult to fit data across the whole infrared range simultaneously
and quantify the errors on each parameter.  Having an analytic
approximation for this functional form (e.g. Equations~\ref{eq4} and
~\ref{eq5}) makes error propagation and multiple-datapoint fitting
much more straightforward.

\subsubsection{Adjusting the 2--4 free parameters}

There are six total parameters in this fit: the greybody normalization
($N_{\rm bb}$), the powerlaw normalization ($N_{\rm pl}$), greybody
temperature ($T$), emissivity index ($\beta$), mid-infrared powerlaw
slope ($\alpha$), and mid-infrared turn-over wavelength
($\lambda_{c}$).  Since the turnover wavelength of the mid-infrared
powerlaw would depend on the turnover point of the cold dust greybody
component and $N_{\rm pl}$ determines the flux scaling of the powerlaw
term relative to the greybody, neither parameter should not be
considered 'free' in the sense of the other four.  Both $\lambda_{c}$
and $N_{\rm pl}$ are tied to the best-fit values of $N_{\rm bb}$, $T$
and $\alpha$ such that the total SED resembles both the
\citet{kovacs10a} powerlaw temperature distribution SED and the
\citet{blain03a} piece-wise matched-gradient SED.

The turnover wavelength, $\lambda_{c}$ is set to $3/4$ the wavelength
where the gradient of the greybody is $\alpha$, as in Blain~\etal.
$\lambda_{c}$ is both a function of $\alpha$, $T$ and the adopted
opacity model, and can be approximated as
$\lambda_{c} = (T\times(a_{1}+a_{2}\alpha))^{-1}$
(optically thin) and 
$\lambda_{c} = ((b_{1}+b_{2}\alpha)^{-2} + (b_{3}+b_{4}\alpha)\times T)^{-1}$
(general opacity).  The values of the coefficients are given in
Table~\ref{tab1}.  Note that the factor of $3/4$ is incorporated so
that the juncture of the powerlaw and greybody is most smooth (i.e. it
does not have a physical interpretation, although is related to the
falloff rate of $e^{-x^2}$).  The coefficient of the powerlaw term,
$N_{\rm pl}$, is tied to the normalization of the greybody term and
solves the condition $N_{\rm
  pl}\lambda_{c}^\alpha=$S$_{\nu}(\lambda_{c})$, where S$_{\nu}$ is
given by Equations~\ref{eq1} or ~\ref{eq3}.  Having fixed these two
parameters reduces the number of free parameters to 2--4.

Depending on the amount of FIR photometric data available, further
constraints can be made to reduce the number of free parameters in the
fit described in Equation~\ref{eq4} or Equation~\ref{eq5} from four
($N_{\rm bb}$, $T$, $\beta$, and $\alpha$) to two ($N_{\rm bb}$ and
$T$). The number of free parameters should never exceed the number of
independent data points minus one.  The emissivity $\beta$ varies from
1-2.5 in the literature for individual sources, although the vast
majority of works assume a fixed value of $\beta$=1.5 \citep[][among
  others]{chapman05a,pope06a,casey09b}.  Varying $\beta$ does not have
a very significant impact on FIR luminosity or dust temperature, but
does have a strong impact on the slope of the Rayleigh-Jeans tail at
rest-frame $\lambda\,\ge$200\um.  If there are $>$3 independent
photometric points at $\lambda\,\ge$200\um, then $\beta$ can be
constrained with the fit; otherwise, the fixed value of $\beta$=1.5 is
suggested.  The mid-infrared powerlaw slope can be constrained
similarly; if $>$3 photometric points are available at rest-frame
$\lambda\,\le$70\um, then $\alpha$ can be measured.  Otherwise, a
fixed value of $\alpha$=2.0 is consistent with most sources and is
directly comparable to the mid-infrared portion of SED templates
presented in the next section.  Typical values of $\alpha$ range from
0.5 (nearly flat, consistent with quite a bit of warm dust) to 5.5
(very steep, consistent with very little warm dust).  Note that values
of $\alpha<$1 lead to a divergent luminosity in the near-infrared, so
some additional cutoff should be placed at short wavelengths to avoid
a nonphysical interpretation.

\subsubsection{Defining L$_{\rm IR}$ and T$_{\rm dust}$}

The normalization of the fits is governed by the infrared luminosity,
L$_{\rm IR}$, whose integration limits have varied throughout the
literature. $L_{\rm IR}$ is intended to represent the bulk of a
galaxy's dust emission.  For galaxies which are very infrared-bright
($L_{\rm IR}>$10$^{11}$\,L$_\odot$), $L_{\rm IR}$ is a proxy for
bolometric luminosity.  $L_{\rm IR}$ is most often taken from
8-1000\,\um\ \citep[e.g.][]{kennicutt98a}. L$_{\rm IR}$ also goes by
different names in the literature: L$_{\rm TIR}$ (``total IR'') or
L$_{\rm FIR}$ (``far IR'') and can be integrated from
3--1100\,\um\ \citep[e.g.][]{chapman03b},
40--120\,\um\ \citep[e.g.][]{younger09a}, or
40--1000\,\um\ \citep[e.g.][]{conley11a}.  The rest of the paper uses
the standard 8--1000\um\ integration limits for the sake of
consistency with most of the literature; however, if only the
cold-dust, star-formation dominated infrared luminosity is desired,
more appropriate limits would be 40--1000\um.

In the initial years of characterizing {\it IRAS}-detected IR-luminous
galaxies, the integrated IR flux was approximated by: 
\begin{equation}
F_{\rm FIR(40-500)}\,=\,1.26\times10^{-14}\,\{2.58\,f_{\rm 60} + f_{\rm 100}\}
\label{eq6}
\end{equation}
in W\,m$^{-2}$ \citep[and where flux densities, $f$, are given in
  $Jy$, see the review in][]{sanders96a}. Including all {\it IRAS}
bands and integrating over a wider wavelength range, 8-1000\um,
this becomes:
\begin{equation}
F_{\rm IR(8-1000)}\,=\,1.8\times10^{-14}\{13.48\,f_{\rm
  12}+5.16\,f_{\rm 25}+2.58\,f_{\rm 60}+f_{\rm 100}\}\, .
\label{eq7}
\end{equation}
Luminosities are then $L_{\rm IR}$=$4\pi\,$D$_{L}^{2}$F$_{\rm IR}$.
Section \ref{s:compare} discusses the accuracy of this widely accepted
luminosity approximation to the \citeauthor{u12a} sample.

Note also that the dust temperature $T$ as given in
Equations~\ref{eq1}--\ref{eq5} represents the intrinsic galaxy
temperature, which is different than the inverse ``peak wavelength''
temperature as measured by Wein's displacement law,
i.e. $\lambda_{max}$=\,$b$/$T_{\rm d}$ (where
$b$=2.898$\times$10$^{3}$\um\,K), which only applies to perfect
blackbodies.  Figure~\ref{fig:tdlpeak} shows the intrinsic dust
temperature $T$ against peak wavelength dust temperature, $T_{d}$ for
the above formulations.  Figure~\ref{fig:tdlpeak} also shows $T_{d}$
against {\it IRAS} color, $log(S_{60}/S_{100})$, for
Equations~\ref{eq1}--\ref{eq5} and the IR template SEDs discussed in the
next section.

Far-infrared color is often taken as a proxy for dust temperature.
Works in the literature vary the use of different fitting methods,
some using IR template fits, some using optically thin greybodies,
some including a warm dust component and some not. The convention of
referring to a single dust temperature is meaningless unless it refers
to the same measurement made with consistent methodology.  However,
the Wein's displacement law value of dust temperature, $T_{\rm d}$,
does not vary greatly between fitting techniques, so it is insensitive to
biases in fitting methods, making it a good diagnostic of the
characteristic dust temperature for the whole system.  Throughout the
rest of the paper, $T_{\rm d}$ is used to represent galaxies' dust
temperatures and this method also permits fair comparisons across
literature sources.

\subsection{IR Template SEDs}\label{s:templates}

Template spectral energy distributions are taken from four sources,
all of which are widely used in the literature to derive values such
as $L_{\rm FIR}$ (thus star formation rate, SFR), T$_{\rm dust}$, and
M$_{\rm dust}$ for nearby and high-$z$ galaxies.  The first 105
template SEDs are from \citet*{chary01a}, whose models are developed
to represent existing data from $\sim$0.4-850\um\ on nearby galaxies
(and which vary as a function of L$_{\rm FIR}$, T$_{\rm dust}$ and PAH
strength).  We also use the 64 phenomenological models of
\citet{dale01a}, supplemented by FIR/submillimeter data in
\citet*{dale02a}.  The SEDs from \citet*{chary01a} and
\citet*{dale02a} are the models most frequently used in the literature
\citep[e.g.][and many others]{chapman05a,chapin09a}.  However, the
radiative transfer model SEDs from \citet{siebenmorgen07a} are also
included, and their parameter space spans much more diverse SED types
(by varying nuclear radius, visual extinction, ratio of OB stellar
luminosity to total, and dust density).  While the
\citet{siebenmorgen07a} work contains $\sim$7000 SEDs, we limit
ourselves to the 120 most luminous SEDs, consistent with ULIRG
infrared luminosities.  Finally, we include 25 of the most IR-luminous
models from the PAH-motivated work of \citet*{draine07a} which have
been used for high-$z$ samples \citep[e.g. see][]{marsden11a}.

The primary observable difference between the infrared template SEDs
in this section and the coupled greybody/powerlaw is the inclusion of
the PAH emission features and silicate absorption features.  Readers
interested in the relation between PAH emission and FIR dust emission
should clearly use template SEDs instead of simple greybody/powerlaw
fits.  However, in terms of the basic physical properties extracted
from FIR SEDs$-$$L_{\rm IR}$, $T_{\rm dust}$, and $M_{\rm
  dust}$, there is not a significant detriment to using the
greybody/powerlaw fitting technique.  The PAH emission lines at
7.7\um, 11.2\um, and 12.8\um\ contribute to the integrated IR
luminosity on the order of 5\%, although their contribution is often
negated by the presence of the 9.7\um\ absorption feature.

The mid-infrared spectra of dusty starbursts have been shown to vary
substantially at both low-$z$ and high-$z$
\citep{brandl06a,pope08a,menendez-delmestre09a}.  Without a direct
mid-infrared spectrum, it is impossible to know whether or not PAHs or
absorption features are contributing significantly to broadband
photometric measurements.  At certain rest-frame wavelengths,
FIR photometric measurements are likely to deviate positively or
negatively from a mid-infrared powerlaw (e.g. at $\sim$8\um\ in the
former case and 10\um\ in the latter case).  In those cases, infrared
template SEDs should be used to more accurately determine the
structure of mid-infrared emission.

\section{SED comparisons with data}\label{s:compare}

Testing and comparing the different IR Template SEDs from
Section~\ref{s:templates} and SED fitting methods from
Section~\ref{s:sedfitting} requires a sample of well-constrained
FIR-bright galaxies with accurate FIR photometric measurements.  To
date, the most extensively imaged IR galaxies$-$besides the handful of
high-$z$ brightly lensed sources \citep[e.g. the ``cosmic
  eyelash,''][]{swinbank10a}$-$are from the local {\it IRAS}-selected
Revised Bright Galaxy Sample \citep[RBGS;][]{sanders03a}, particularly
the subset which form part of the Great-Origins All Sky LIRG Survey
\citep[GOALS;][]{armus09a}.  The revised standard-aperture photometry
for 65 galaxies in the GOALS sample is summarized and presented in
\citet{u12a}, and used herein at rest-frame wavelengths 5--2000\um.

\begin{table}
\centering
\begin{tabular}{l@{ }l@{ }r}
\hline\hline
\multicolumn{3}{c}{
$S(\lambda) =
N_{\rm bb}\,\frac{(1-e^{-(\frac{\lambda_{0}}{\lambda})^{\beta}})(\frac{c}{\lambda})^{3}}{e^{hc/\lambda\!kT}-1}
+
N_{\rm pl}\,\lambda^{\alpha}\,e^{-(\frac{\lambda}{\lambda_{c}})^2}$} \\
\hline
Free Parameters: & \multicolumn{2}{l}{N$_{\rm bb}$, $T$, ($\beta$, $\alpha$)} \\
\hline
Emissivity & $\beta$ & 1.60$\pm$0.38 \\
Mid-IR Powerlaw Slope & $\alpha$ & 2.0$\pm$0.5 \\
Wavelength where opacity is unity & $\lambda_{0}$ & $\equiv$200\,\um \\
Powerlaw turnover wavelength & $\lambda_{c}$ & $\equiv$3/4 L($\alpha$,$T$) \\
\multicolumn{3}{r}{where L($\alpha$,$T$) = (($b_{1}$+$b_{2}\,\alpha$)$^{-2}$ + ($b_{3}$+$b_{4}\,\alpha$)$\times\,T$)$^{-1}$} \\
 & $b_{1}$ & 26.68 \\
 & $b_{2}$ & 6.246 \\
 & $b_{3}$ & 1.905$\times$10$^{-4}$ \\
 & $b_{4}$ & 7.243$\times$10$^{-5}$ \\
Normalization of powerlaw term & $N_{\rm pl}$ & \\
\multicolumn{3}{r}{$\equiv\,N_{\rm bb}\,\frac{(1-e^{-(\lambda_{0}/\lambda_{c})^{\beta}})\lambda_{c}^{-3}}{e^{hc/\lambda_{c}kT}-1}$} \\
Characteristic Dust Temperature ($\ne\,T$) & T$_{\rm d}$ & 1/$\lambda_{\rm peak}$ \\
\hline\hline
\end{tabular}
\caption{Adopted Best-Fit SED fitting equation and measured parameters
  based on the GOALS sample.  The measurement of $\beta$ is based on
  48 out of the 65 galaxies which have 850--1.1\,mm data, while the
  measurement of $\alpha$ is based on the full sample of 65 (U)LIRGs.}
\label{tab1}
\end{table}

Modified greybodies plus mid-infrared powerlaw SEDs are fitted to data
as described in Section~\ref{s:sedfitting}, specifically to
Equation~\ref{eq4}, taking both flux density uncertainties and
non-detection upper limits into account\footnote{ An IDL function {\tt
    cmcirsed.pro}, which can be used to fit an SED of the form in
  Eq~\ref{eq4} or Eq~\ref{eq5} to real data, is publicly available at
  {\tt www.ifa.hawaii.edu/$\sim$cmcasey/research.html}.}.  The
general-opacity fits are chosen for this local sample over the
optically-thin fits due to a subtle difference in SED shape
around rest-frame wavelengths $\sim$20-40\um.

For sources with more than three long-wavelength data points, 
$\beta$ is kept as a free parameter; otherwise it is fixed to
$\beta$=1.5.  Mid-infrared powerlaw slope $\alpha$ is allowed to vary
since the number of short-wavelength points ($<$70\um) is always more
than three.  The range of $\alpha$ values is found to vary between
$\alpha$=0.5--5.5. Dust temperature and luminosity are left to vary.
Template IR SEDs from Section~\ref{s:templates} are fitted to data
with a $\chi^{2}$ minimization method, where the normalization
(i.e. N$_{\rm bb}\propto$L$_{\rm IR}$) is a free parameter.

\begin{figure*}
\centering
\includegraphics[width=2.0\columnwidth]{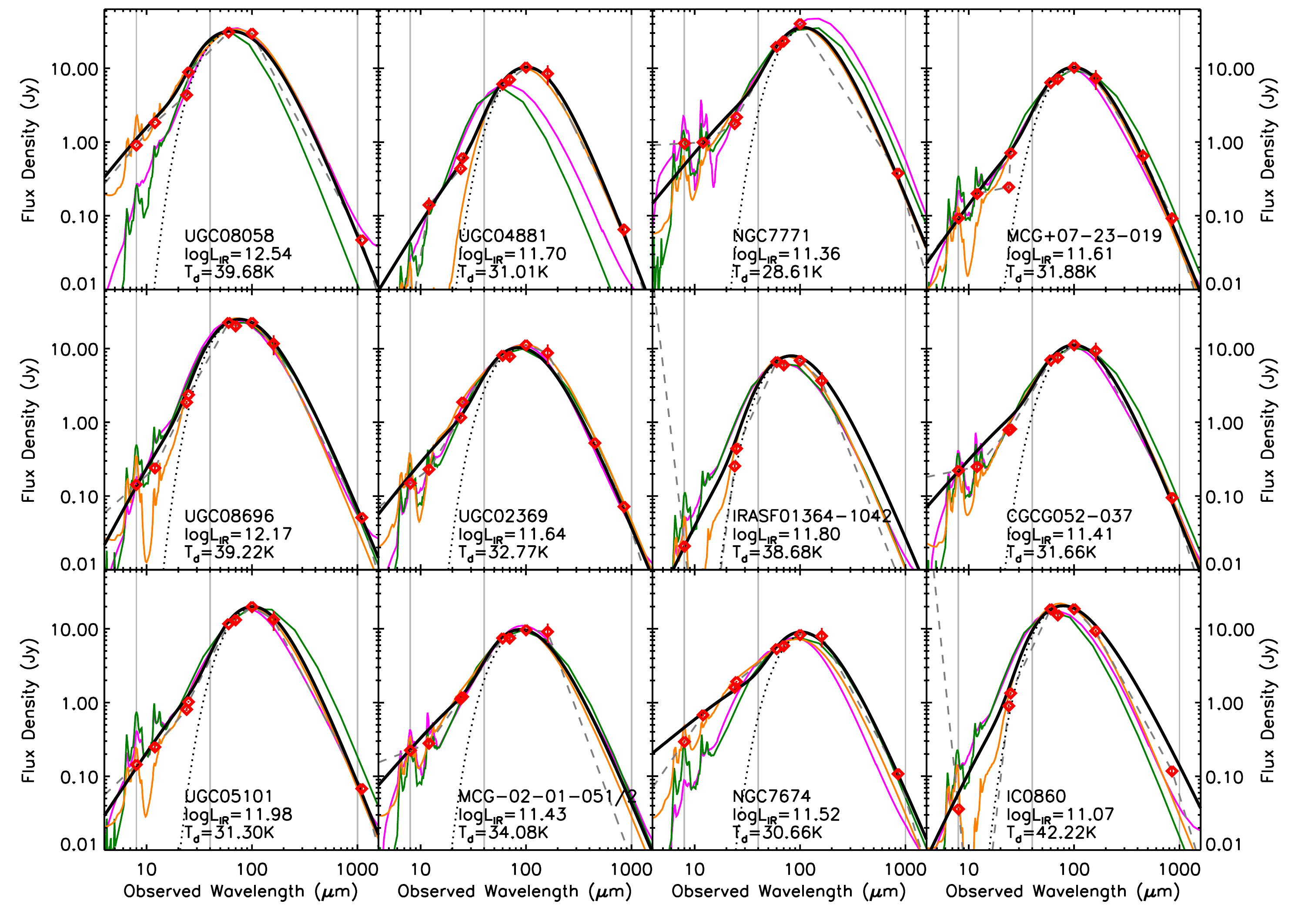}
\caption{ Example best-fit SEDs to a subset of the GOALS sample, whose
  photometry is presented in \citet{u12a}.  Photometric data points
  are shown as red diamonds with uncertainties illustrated (sometimes
  smaller than the size of the data point).  The best-fit SED fit from
  Equation~\ref{eq4} is shown in black, with the underlying greybody
  distribution shown as a dotted black line.  The best-fit
  \citeauthor*{chary01a} template SED is shown in magenta,
  \citeauthor*{dale02a} SED shown in green, and
  \citeauthor*{siebenmorgen07a} SED shown in orange.  A ``dummy SED''
  is also constructed which linearly interpolates between data points
  in log$\lambda$-log$S_{\nu}$ space (dashed gray line).  The
  8--1000\um\ and 40--1000\um\ integration limits are marked as light
  gray vertical lines for reference.
}
\label{fig:example}
\end{figure*}

Figure~\ref{fig:example} shows several randomly-selected examples of
best-fit SEDs for GOALS sources.  The various best-fit IR templates
are colored, while the fitted SED is shown in black.  The underlying
greybody function to the fitted SED is shown as a dotted line.  Since
$>$8 FIR photometric points exist for the GOALS sample, a dummy SED
can also be constructed by linearly extrapolating between photometric
points (shown in gray).  While this is a rough approximation, the
advantage of this extrapolation is that this dummy SED makes no
intrinsic assumptions as to SED shape.  The subset of sources shown in
Figure~\ref{fig:example} represents a wide variety of templates and
fitted SEDs and is representative of the fits to the whole
\citet{u12a} sample.

\begin{figure}
\centering
\includegraphics[width=1.0\columnwidth]{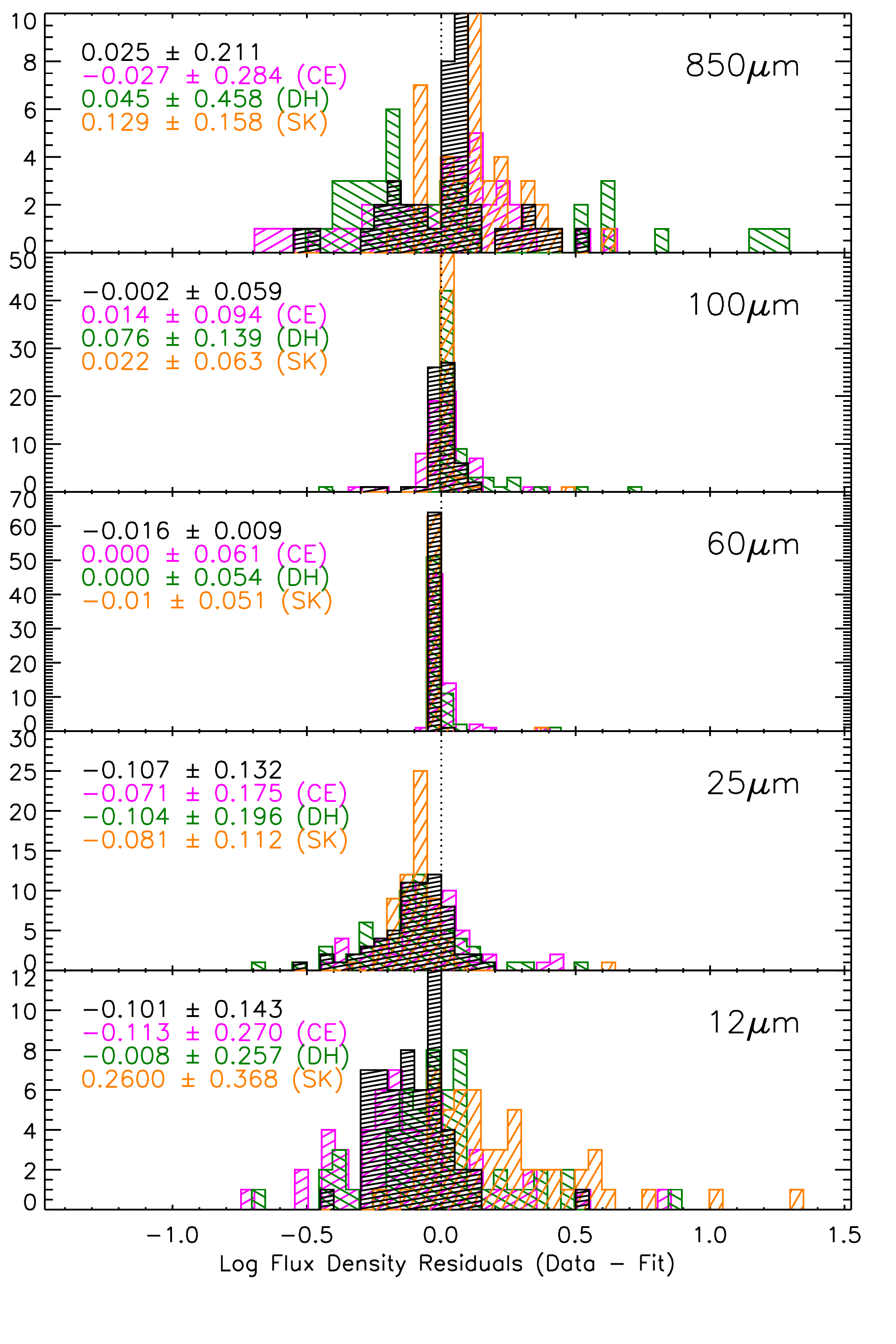}
\caption{ The residuals of the fits to the GOALS sample between
  template libraries and the coupled greybody/powerlaw fits for all 65
  galaxies, for the greybody/powerlaw fits (black), Chary \&\ Elbaz
  fits (magenta), Dale \&\ Helou fits (green) and Siebenmorgen
  \&\ Kr\"{u}gel fits (orange). The mean and standard deviations of the
  residual distributions for each fitting method are given in the upper
  left-hand corner.  The residuals are lowest for greybody/powerlaw
  fits, followed by the Siebenmorgen \&\ Kr\"{u}gel fits.}
\label{fig:residuals}
\end{figure}

The residuals of the fits with respect to data are shown in
Figure~\ref{fig:residuals}.  The residual is computed as the
difference in log$S_{\nu}$ between the data points and the best fit
SED at each of the following wavelengths: 12\um, 25\um, 60\um, 100\um,
and 850\um.  Figure~\ref{fig:residuals} shows the distribution in the
residuals at each wavelength for the \citeauthor*{u12a} GOALS sample
for each fit.  Black represents the residuals from the fitting method
described in Section~\ref{s:sedfitting}, and the residuals of the IR
templates are colored as in Figure~\ref{fig:example}.  At each
wavelength, the mean and standard deviation of the residual
distribution is given in the top left.

At \emph{all} wavelengths, the fitted SED from Equation~\ref{eq4} is
statistically a better fit to the data than any IR templates, despite
having fewer free parameters than the SED templates; a summary of
derived quantities from this fit is given in Table~\ref{tab1}.
However, some templates are more accurate than others; the
\citeauthor*{siebenmorgen07a} templates provide the best fit of the
templates, significantly better than both \citeauthor*{chary01a} and
\citeauthor*{dale02a} templates.  However, all seem to suffer at short
wavelengths where there is naturally no flexibility built into the
models.  This demonstrates that galaxies' mid-infrared properties are
not tightly correlated with their far-infrared properties, as perhaps
previously thought \citep[since 24\um\ is often used to infer $L_{\rm IR}$,
  e.g.][on {\it Spitzer} samples, among many others]{le-floch05a}.

Another reason the \citeauthor*{siebenmorgen07a} templates have lower
residuals than the other two template libraries is that they contain more
templates with which to compare the data.  Even at the highest
luminosity end ($L_{\rm IR}>$10$^{11}$L$_\odot$), there are still 120
template models in the \citet*{siebenmorgen07a} library.  In contrast,
the \citeauthor*{chary01a} library contains 105 template SEDs and
\citeauthor*{dale02a} contains 64 template SEDs.

This direct comparison between SED fitting and SED templates
demonstrates that using IR templates will always provide a less
accurate fit than simple greybody/powerlaw fitting.  This is
not meant to suggest that IR template SEDs are not of great use; on
the contrary, they provide very good constraints on the relationships
between FIR emission and PAH emission, and sources' dust composition,
which simple greybody fitting cannot do.  However, if the purpose of
FIR SED fitting is to measure a source's FIR luminosity, dust
temperature and dust mass accurately, this work demonstrates that
direct fitting, as in Section~\ref{s:sedfitting}, is best.

\subsection{Interpretation of best-fit parameters}

The \citet{u12a} sample has been used to constrain the mean SED
parameters $\beta$, the spectral emissivity index, and $\alpha$, the
mid-infrared powerlaw slope.  How do we interpret the measurements
$\beta$=1.60$\pm$0.38 and $\alpha$=2.0$\pm$0.5?  Here I draw on the
detailed discussions in \citet*{scoville76a}.

The measured value of the emissivity index, $\beta$, is spot on its
presumed value of $\approx$1.5.  Although not new information, this
suggests that even the cold dust component (dominating at
$\lambda$$>$50\um) has a temperature gradient whereby the dust
farthest from the luminous starburst is slightly colder, which
deferentially boosts the flux density at the longest wavelengths.
This boosting (above a simple blackbody) becomes enhanced when the
density distribution falls off slowly, and with a steep density drop
off, $\beta$ should increase, and the far-infrared ``slope'' should
increase.

The mid-infrared powerlaw slope, $\alpha$, probes the warmer,
more-compact dust, and can be used to estimate the radial density
profile.  Our measurement of $\alpha$=2.0$\pm$0.5 agrees with earlier
findings of $\alpha=$1.7--2.2 from analysis of IRAS and distant
galaxies in \citet{blain02a}.  Assuming there are sufficient hot dust
grains near the interior and the dust is optically thin,
$S_{\lambda}\propto\lambda^{\alpha}$ where $\alpha$=2.0$\pm$0.5
translates to a density profile of $r^{-0.5\pm0.2}$, consistent with
$r^{-1/2}$.  In other words the dust density profile is relatively
flat and perhaps indicative of the diffuse nature of the dust around
these starbursts, their distributions perhaps originating from the
violent interacting nature within the galaxies.

\section{Derived Quantities}\label{s:derived}

Section~\ref{s:compare} used the GOALS sample \citep{armus09a,u12a} to
demonstrate that direct greybody/powerlaw SED fitting, as described in
Section~\ref{s:sedfitting}, is a more accurate fit to data than
best-fit IR SED template libraries from the literature.  In this
section, the derived quantities $L_{\rm IR}$ (luminosity), $T_{\rm d}$
(dust temperature) and $M_{\rm d}$ (dust mass) are compared to see
what can be expected using different SED fitting methods.

\subsection{IR Luminosity and Star Formation Rates}

The IR luminosity is most often integrated in the range 8-1000\um,
i.e. $L_{\rm
  IR}\,=\,4\pi\,D_{L}^{2}\int_{\lambda=8\mu\!m}^{1000}\,S_{\nu}d\nu$.
The range 8-1000\um\ encompasses some of the most prominent PAH
emission features in addition to the greybody emission peak.  The
boundaries are still somewhat arbitrarily drawn, since 8\um\ sits
awkwardly on top of the 7.7\um\ PAH emission feature, and
1000\um\ splits the cold-dust greybody neither at its turnover point,
where radio synchrotron emission begins to dominate, nor near the
peak.  On average, the cold-dust greybody component$-$thought to be
exclusively heated by star-formation processes$-$comprises
74$\pm$11\%\ of $L_{\rm IR(8-1000)}$, while warm AGN-heated dust and
PAH emission components comprise 26$\pm$11\%.

\begin{figure}
\centering
\includegraphics[width=1.0\columnwidth]{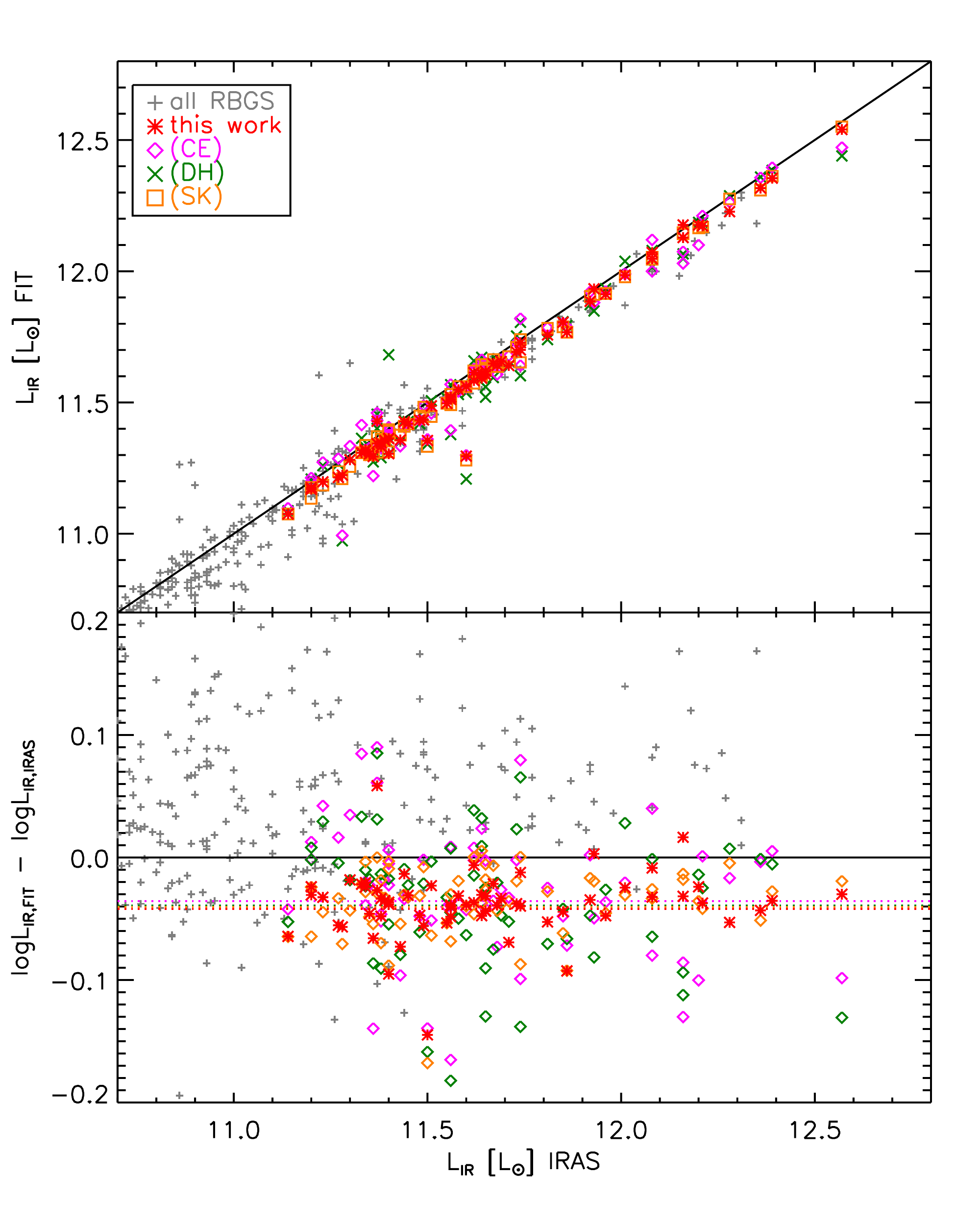}
\caption{ The fitted RBGS IR luminosities, ``{\it IRAS}
  luminosities,'' are measured as a linear combination of {\it IRAS}
  12-100\um\ flux densities, see Eq~\ref{eq7} or \citet*{sanders96a}.
  Here {\it IRAS} luminosities are plotted against fitted luminosities
  from the direct fitting method (Section~\ref{s:sedfitting}), and
  template SED libraries.  The full RBGS sample of 625 sources from 
  \citet{sanders03a} is included, and luminosities refitted using {\it IRAS}
  photometry alone and Eq~\ref{eq4}, fixing $\beta$=1.5 (scatter is
  larger for this sample due to poorer photometric constraints).  At
  bottom are the residuals, luminosity against the difference in
  logL$_{\rm IR}$.  All template libraries and fitted SEDs demonstrate
  that {\it IRAS} luminosities are overestimated by $\sim$0.04\,dex,
  or $\sim$10\%.}
\label{fig:luminosity-check}
\end{figure}
Figure~\ref{fig:luminosity-check} shows the 8--1000\um\ IR
luminosities of the \citeauthor{u12a} sample and luminosity residuals
when contrasting methods.  The $x$-axis is the IR luminosity as
inferred by Equation~\ref{eq7}, also as summarized in
\citet{sanders96a} and \citet{armus09a}.  The whole RBGS sample is
also included, refitting all luminosities using the method from
Section~\ref{s:sedfitting} (due to limited long-wavelength
photometry, $\beta$=1.5 is fixed).  Similarly, RBGS luminosities are
compared to template SED-inferred luminosities.  Both the fitted SEDs
and template SEDs have lower IR luminosities than predicted using the
RBGS Equation (Eq~\ref{eq7}) by $\sim$0.02-0.04\,dex.  Both
\citeauthor*{chary01a} and \citeauthor*{dale02a} templates are
0.02$\pm$0.07\,dex lower in luminosity than {\it IRAS} predictions,
while \citeauthor*{siebenmorgen07a} templates are 0.03$\pm$0.05\,dex
lower, and fitted SEDs are 0.04$\pm$0.04\,dex lower.  These
differences translate to luminosity overpredictions from RBGS of 10\%,
thus overestimated SFRs by 10\%.

Note that the \citet{kennicutt98a} scaling relation between $L_{\rm
  IR(8-1000)}$ and SFR take both the contribution of ``infrared
circus'' cold-dust emission and ``warm'' dust around young star
forming regions into account and is generated using stellar synthesis
models \citep{leitherer95a} for continuous bursts aged 10-100\,Myr.
Since the Kennicutt scaling is not determined as an empirical relation
between, e.g. H$\alpha$ SFR and L$_{\rm IR}$, there is no need to
recalibrate it to correct for the overestimated L$_{\rm IR}$; however,
the FIR star formation rates of the local (U)LIRG sample should be
revised to reflect the systematic offset.


Regardless of the systematic offset, infrared-based star formation
rates are only accurate to $\sim$10\%\ mostly due to the variation in
mid-infrared properties, likely not directly scaling to the galaxies'
star formation rates, since complex processes dominate the
mid-infrared.  Similarly, slight changes in any SED which impacts the
mid-infrared (e.g. changes in opacity assumptions, assumed $\alpha$,
PAH contributions) will impact the derived star formation rate.  For this
reason it is worth emphasizing that the systematic offset is of minor
significance.  Also worth pointing out is that the range
8--1000\um\ is not ideal to generate precise star formation rates.
\begin{figure}
\centering
\includegraphics[width=0.9\columnwidth]{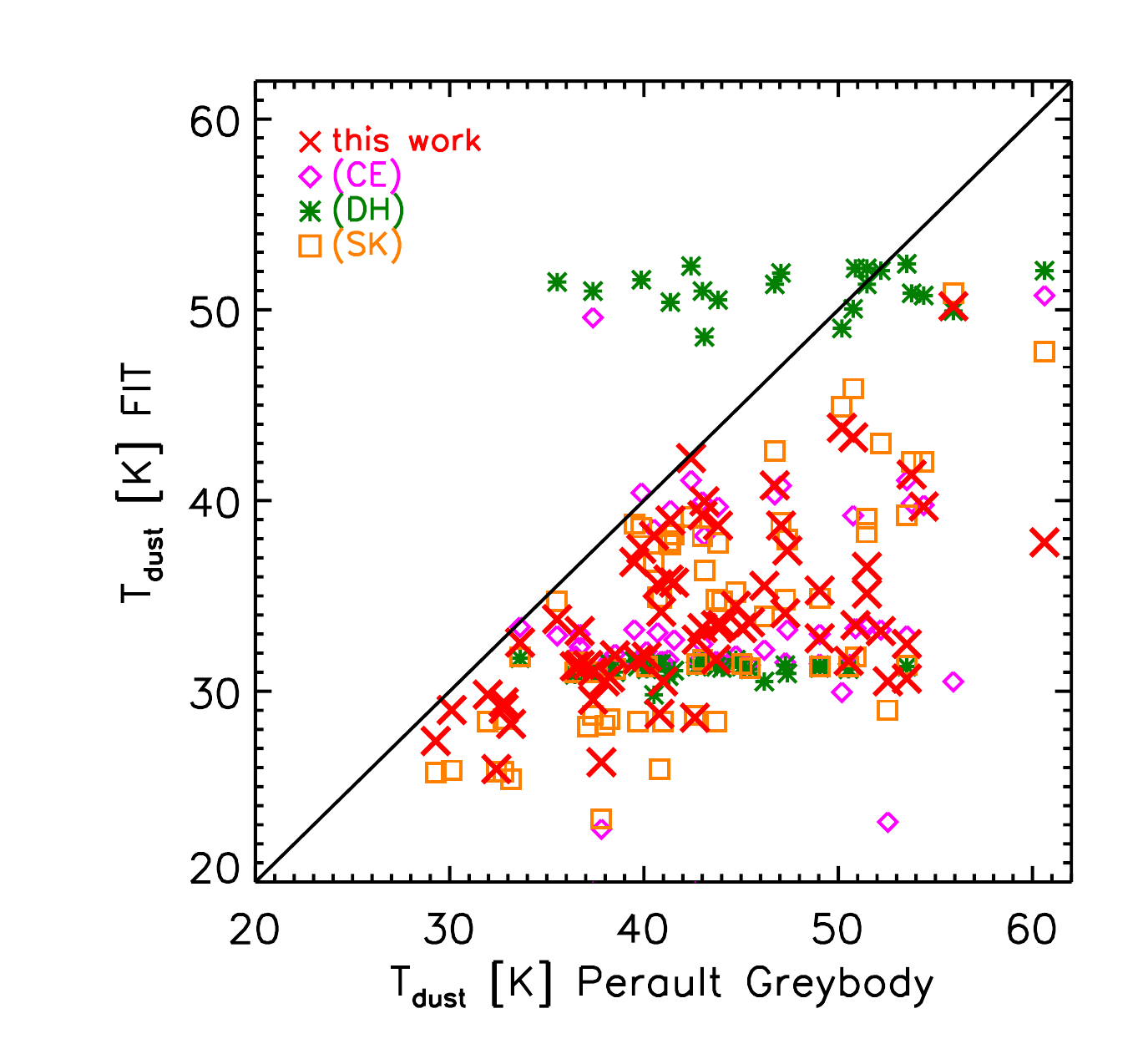}
\caption{ A comparison of dust temperatures from Perault\,(1987),
  described also in \citet*{sanders96a}, against dust temperatures
  found using methods in Section~\ref{s:sedfitting} and from template
  SEDs.  Colors and symbols are as in
  Figure~\ref{fig:luminosity-check}.  In all cases, a single greybody
  fit of the form described in Eq~\ref{eq1} or \ref{eq3} (which was
  the method used in Perault\, 1987) overestimates dust temperature by
  10\,K on average.  Caution is necessary when using dust
  temperatures from template SEDs since they are quantized.
}
\label{fig:td-check}
\end{figure}

Ideally, one would extract the integrated luminosity from the
cold-dust greybody component over all wavelengths where it dominates.
This can be estimated precisely if the fitting method in
Section~\ref{s:sedfitting} is used, as the greybody component can then
be constructed independently after fitting to a joint greybody and
powerlaw.  However, to make fair comparisons with different SED
fitting methods, the optimal alternative limits of integration should
be taken as 40--1000\um\ as in \citet{conley11a}.  Longward of 40\um,
the cold-dust greybody dominates at all dust temperatures $<$100\,K.

\subsection{Dust Temperature}

Comparing dust temperatures between models first requires an
understanding of the dust temperature convention for the original work
on the \citet{u12a} GOALS sample objects.  Dust temperatures were
originally fitted in Perault~(1987; PhD Thesis) by fitting a single
temperature dust emissivity model ($\epsilon\propto\nu^{-1}$) to the
flux in all four {\it IRAS} bands \citep[briefly described
  in][]{sanders96a}.  In other words, these dust temperatures are
fitted to data ranging 12-100\um\ to either Eq~\ref{eq1} or
Eq~\ref{eq3}.  Assumptions made with regard to opacity are not stated;
however, the impact on peak dust temperature is minimal.  It could be
that the dust temperature reported in Perault~(1987) is indeed $T$
taken from e.g. Eq~\ref{eq3} rather than ``peak wavelength'' dust
temperature $T_{\rm d}$, although it is unclear.  Regardless,
refitting greybodies of either form specified in Eq~\ref{eq1} or
~\ref{eq3} to only the four {\it IRAS} datapoints causes the dust
temperatures $T_{\rm d}$ to drop by $\sim$10\,K.  This is shown in
Figure~\ref{fig:td-check} (and was illustrated clearly in
Figure~\ref{fig:schematic}, panels A and B).  The difference between
dust temperatures of simple greybodies and greybodies which
incorporate a mid-infrared component has been known \citep[see also
  the recent detailed discussion in][]{hayward12a}, yet few works
are explicitly clear on how their dust temperature measurements should
be interpreted.

The Perault/RBGS dust temperatures are also found to be significantly
hotter than dust temperatures measured from the SED template
libraries.  However, any dust temperatures extracted from the template
SEDs should be used with great caution, as the dust temperatures are
quantized in all cases (as was seen at right in
Figure~\ref{fig:tdlpeak}).  The \citet*{dale02a} templates exhibit the
poorest constraints on dust temperature, as the templates only have
two dust temperatures: $T_{\rm d}\approx$31\,K and $T_{\rm
  d}\approx$51\,K.  The \citet*{chary01a} templates have four
different dust temperatures, at $\approx$23\,K, 32\,K, 41\,K and
50\,K.  The \citet*{siebenmorgen07a} templates are quantized on yet
finer scales, at 5\,K increments between $\sim$24\,K and 85\,K;
however, the fitted SEDs, as an analytic fit, are not quantized in
$T_{\rm d}$.  For this reason, any dust temperature-based results
should be based on fits of the type described in
Section~\ref{s:sedfitting}, not on template SEDs in the literature.

\subsection{Dust Mass}
\begin{figure}
\centering
\includegraphics[width=0.9\columnwidth]{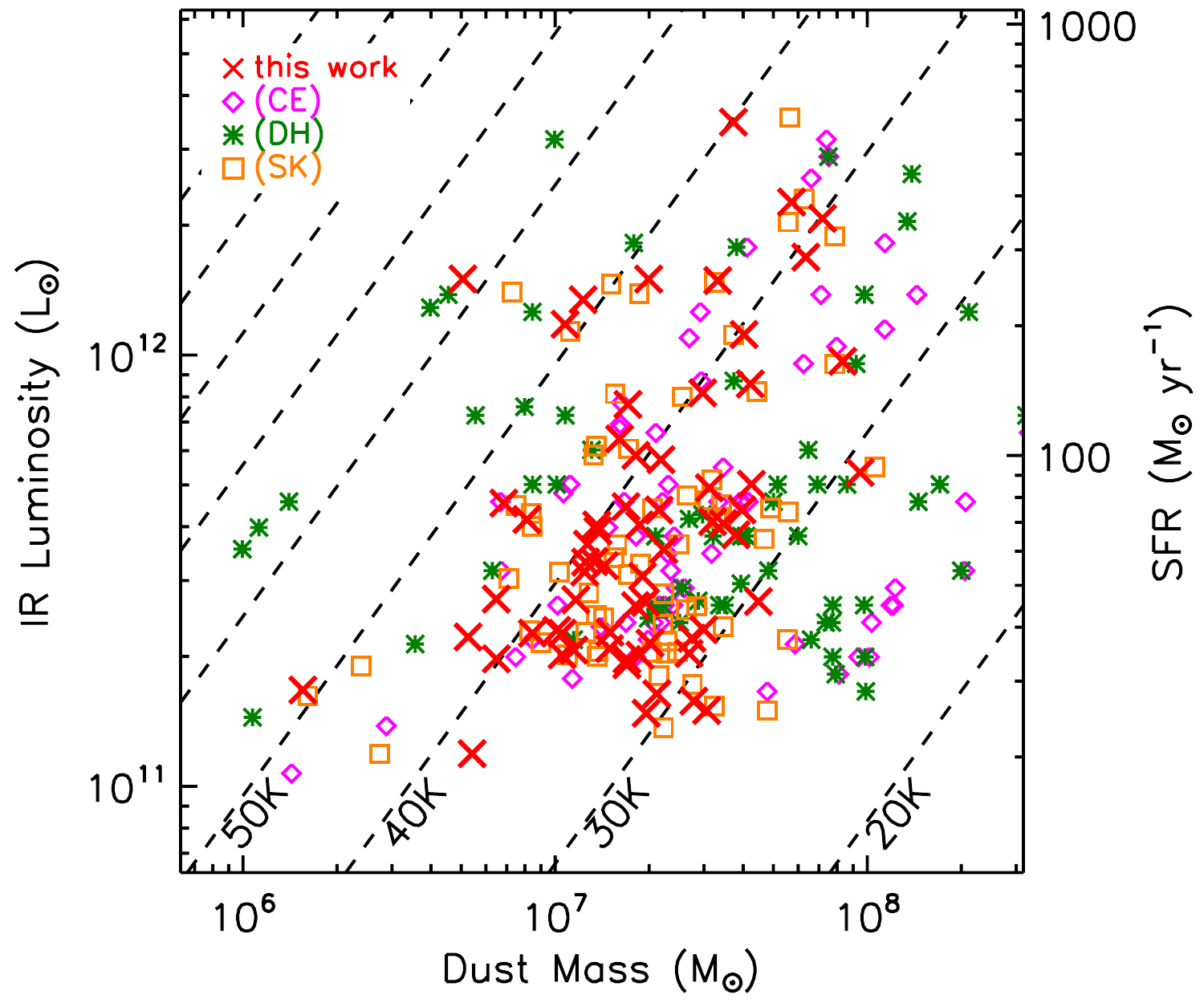}
\caption{ Dust mass against 8-1000\um\ IR luminosity and star
  formation rate (SFR) measured from the infrared.  Isotherms are
  marked with dashed lines increasing by 10\,K.  This emphasizes that
  (a) dust mass is incredibly sensitive to dust temperature and that
  dust mass should only be computed when dust temperature is
  adequately constrained, and (b) the lack of correlation between
  integrated IR luminosity and the Rayleigh-Jeans cold-dust component
  of IR SEDs (the best probe of dust mass).  The latter point points
  out that using direct scaling relations between, e.g. $S_{850}$ and
  SFR is not recommended.
%
}
\label{fig:mdust-lfir}
\end{figure}

Dust mass is related to IR flux density and dust temperature via
\begin{equation}
S_{\nu}\,=\,\kappa_{\nu}\,B_{\nu}(T)\,M_{\rm d}\,D_{\rm L}^{-2}
\label{eq8}
\end{equation}
where S$_{\nu}$ is the flux density at frequency $\nu$, $\kappa_{\nu}$
is the dust mass absorption coefficient at $\nu$, B$_{\nu}$($T$) is
the Planck function at temperature $T$, M$_{\rm d}$ is the total dust
mass, and D$_{\rm L}$ is the luminosity distance.  The fact that our
sources are not perfect black bodies is accounted for by the dust mass
coefficient $\kappa_{\nu}$ so that the greybody is effectively
represented by the product $\kappa_{\nu}\,B_{\nu}$($T$), and the
sources' luminosity at frequency $\nu$ scales as
$S_{\nu}$/$B_{\nu}$($T$)$\propto\,\nu^{-2}$.  The dust mass is then
\begin{equation}
M_{\rm d}\,=\,\frac{S_{\nu}\,D_{\rm L}^{2}}{\kappa_{\nu}\,B_{\nu}(T)}\,\propto\,\frac{D_{\rm L}^{2}}{\kappa_{\nu}\,\nu^{2}}.
\label{eq9}
\end{equation}
As \citet{draine07b} point out, care should be taken when computing
dust masses using measured dust temperatures since the thermal
emission per unit dust mass at $\lambda$ is
$\propto$($e^{hc/\lambda\!kT}-1$)$^{-1}$.  At wavelengths
$\lambda\le$360\um, dust temperature has a profound effect on dust
mass, since B$_{\nu}$($T$) is very dependent on $T$ (so that a
4\,K difference between 18 and 22\,K results in a 150\%\ increase in
M$_{\rm d}$).  At $\lambda\ge$450\um, M$_{\rm d}$ is less sensitive to
dust temperature.  Note that the dust temperature used in Eq~\ref{eq9}
is the $T$ from Eq~\ref{eq4} and not the ``peak wavelength'' dust
temperature $T_{\rm d}$ characterizing the system as a whole.

Due to poor constraints on dust absorption coefficients and no data
points longward of $\sim$450\,\um, dust masses for the GOALS sample
were not reported until the work of \citet{u12a}, who use the method
described in this paper, Eq~\ref{eq9}, to calculate dust masses at
850\um\ using a dust absorption coefficient of $\kappa_{\rm
  850}$\,=\,0.15\,m$^{2}$\,kg$^{-1}$ \citep{weingartner01a,dunne03a}.
Figure~\ref{fig:mdust-lfir} plots the galaxies' dust masses against IR
luminosity, which highlights two important facts.  First, the
calculation of dust mass is very sensitive to dust temperature,
$M_{\rm d}\,\propto\,L_{\rm IR}\,T^{-5}$.  This highlights that
dust mass should {\it only} be estimated when dust temperature is
adequately constrained.  The second noticeable detail of
Figure~\ref{fig:mdust-lfir} is the lack of correlation between the two
physical properties, $L_{\rm IR}$ and $M_{\rm d}$.  The lack of
correlation is a testament to how large amounts of dust can be formed
quite quickly in starburst galaxies (in contrast, star-forming
galaxies with much more moderate star formation rates overall have
dust masses several orders of magnitude smaller,
$<$10$^{5-6}$M$_{\odot}$).  This also points out that scaling
relations where star formation rate is inferred directly from a long
wavelength flux density measurement, e.g. $S_{\rm 850}$, without good
constraints on $L_{\rm IR}$ or $T$, are highly uncertain.


\subsection{Future Improvements}

The next decade will see great improvements in our understanding of
dust emission in galaxies thanks to vast repositories of data from
large field-of-view mapping bolometers and efficient follow-up with
interferometric imaging.  Facilities less limited by confusion noise
(e.g. {\sc Scuba-2} 450\um\ mapping on JCMT, and future
200-850\um\ data from the Cornell-Caltech Atacama Telescope, CCAT)
will make infrared photometric measurements more accurate, precise and
more numerous.  Submillimeter spectrometers \citep[e.g. like {\sc
    Z-Spec} operated at the Caltech Submillimeter Observatory and
  Atacama Pathfinder EXperiment, e.g. see][ also those being designed
  for CCAT]{bradford09a} will measure {\it spectra} for individual
sources in the infrared and submillimeter, providing crucial
constraints on emissivity while also confirming redshifts through CO
lines and assessing their contribution to continuum photometry
\citep[e.g. see][]{smail11a}.

With this superb data, infrared SED fits can be significantly refined.
Better photometric constraints mean that more free parameters can be
reintroduced into SED fitting, providing a more concrete physical
context.  What does the value of $\beta$ imply for the distribution of
cold dust? What does the value of $\alpha$ imply for the radial dust
distribution?  Is there a wavelength regime where the assumption of a
powerlaw distribution of warm dust temperatures does not hold, and is
that in turn related to the dust distribution or presence/lack of a
luminous AGN?  Are any galaxies better fit with two independent
greybodies rather than one, and does this indicate two distinct dust
reservoirs not yet mixed?

Furthermore, interferometers like ALMA will make it routinely possible
to resolve dust emission on kpc scales within distant galaxies
\citep[as the Submillimeter Array, SMA, has done for some local
  ULIRGs;][]{wilson08a}.  Spatial mapping of the dust distribution in
multiple infrared channels can then be used to assess the mid-IR
powerlaw inferred dust distribution.  Combined with resolved molecular
gas maps obtained through CO emission, routine tests of the
Schmidt-Kennicutt relation \citep{schmidt59a,kennicutt98a} can be
performed on a variety of galaxy types at different stages of their
star formation histories.

A further refinement which can be made to SED fitting described in
this paper is the introduction of PAH emission and Si absorption
modeling on top of the suggested dust continuum.  Given the wide
variety of mid-infrared properties for infrared starbursts, this
modeling is best fit independently from the mid-infrared powerlaw dust
continuum.  This decoupling of the two will be possible with deep
near- and mid-IR data from the James Webb Space Telescope (JWST).

\section{Discussion \&\ Conclusions}\label{s:conclude}

The characterization of infrared-bright galaxies has become
increasingly important in recent years with the introduction of new
mid- to far-infrared observing facilities, including {\it Herschel
  Space Observatory} and the Atacama Large Millimeter Array (ALMA).
Infrared SED fitting techniques vary widely in the literature, from
simple greybody fits to detailed dust emission modeling
templates.

Section~\ref{s:sedfitting} has presented an SED fitting technique
which can be used to fit a wide range of IR data, from sources which
have only 3 IR photometric points to sources with $>$10 photometric
points.  These SED fits do not account for PAH emission in the
mid-infrared, although they produce accurate estimates to a source's
integrated IR luminosity, dust temperature and dust mass.  The fitting
technique is based on a single dust temperature greybody fit linked to
a mid-infrared powerlaw, fit simultaneously to data across
$\sim$5-2000\um.  Without inclusion of the mid-infrared powerlaw, dust
temperatures are overestimated and the short-wavelength data
($<$50\um) fit is likely very poor.  From the measurement of the mean
mid-infrared powerlaw slope $\alpha$, the dust radial density profile
in most local (U)LIRGs is fairly shallow, $\approx\,r^{-1/2}$. IDL
code for the fitting procedure has been made publicly available.

This SED fitting procedure is contrasted with fits to infrared
template libraries generated through dust-grain modeling, including
those of \citet*{chary01a}, \citet*{dale02a} and
\citet*{siebenmorgen07a}.  SED fit quality is categorized by goodness
of fit to the GOALS local LIRG and ULIRG sample photometry reported in
\citet{u12a}.  Since the GOALS sources are the most extensively
surveyed infrared galaxies to date, they provide a good testbed for
SED fitting reliability.  The SED templates and original formulation
of luminosity for the RBGS sample do not fit the data as well as the
fitting method described in section~\ref{s:sedfitting}, leading to
overestimated IR luminosities by 10\%.  Similarly, dust temperatures
estimated for the original RBGS sample \citep{sanders03a} were
overestimated by $\sim$10\,K, due to the different fitting method used
(single greybody versus single greybody plus mid-IR powerlaw).  Dust
temperatures cannot be well constrained with SED templates because
they are quantized and only peak at certain wavelengths corresponding
to fixed dust temperatures.  Clarification is offered on the
calculation of dust mass, its dependence on dust temperature, and the
lack of correlation between dust mass and IR luminosity.

This comparative study should be useful to highlight some of the
current problems facing FIR SED fitting techniques, and the details
often overlooked in ambitious analyses of infrared-galaxy population
trends for galaxies both near and far.

\section*{Acknowledgements}

The author is grateful for support from a Hubble Fellowship provided
by Space Telescope Science Institute, grant HST-HF-51268.01-A.  This
work would not have been possible without the many constructive
conversations about SED fitting and dust properties with Nick
Scoville, Vivian U, Alex Conley, Ed Chapin, and Jonathan Williams.
Special thanks to Ezequiel Treister, Vivian U and Dave Sanders for
inspiring this comparative study, and many thanks to the anonymous
reviewer who offered very constructive comments improving the paper.

\bibliography{caitlin-bibdesk}
\label{lastpage}

\end{document}